\documentclass[11pt]{article}
\usepackage{graphicx}

\begin{document}

\renewcommand{\to}{\rightarrow} 


\begin{titlepage}

\begin{flushright}
DFAQ-TH/2000-04 \\ 
hep-ph/0009290 \\ 
September 2000 
\end{flushright}
\vspace{1.0cm}

\begin{center}

{\LARGE \bf Strong CP problem and mirror world: \\
the Weinberg-Wilczek axion revisited  } 

\end{center}

\vspace{0.3cm}

\begin{center}
{\large 
Zurab Berezhiani,$^{a,b,}$\footnote{E-mail: berezhiani@fe.infn.it,
berezhiani@aquila.infn.it } 
Leonida Gianfagna$^{a,}$\footnote{E-mail: gianfagna@aquila.infn.it} 
and Maurizio Giannotti$^{a,}$\footnote{
E-mail: giannotti@aquila.infn.it } 
} \\
\end{center}

\begin{center}
$^a$ {\large \it Dipartimento di Fisica, Universit\`a di L'Aquila, I-67010
Coppito, AQ, and \\ 
INFN, Laboratori Nazionali del Gran Sasso,
I-67010 Assergi, AQ, Italy } \\
$^b$ {\large \it Andronikashvili Institute of Physics, 
Georgian Academy of Sciences, \\
 38077 Tbilisi, Georgia } 
\end{center}

\vspace{5mm} 

\begin{abstract}

A new possibility for solving the strong CP-problem is 
suggested. It is based on the concept of a mirror world 
of particles, with the gauge symmetry and Lagrangian 
completely  identical to that of the observable particles.  
We assume that the ordinary and mirror sectors share 
the same Peccei-Quinn symmetry realized {\it \`a la} 
Weinberg-Wilczek, so that the $\theta$-terms are 
simultaneously canceled by the axion VEV in both worlds. 
This property remains valid even if the symmetry between 
two sectors is spontaneously broken and the 
the weak scale of the mirror world  
is larger than the ordinary weak scale, 
in which case also the mirror QCD scale 
becomes larger than the ordinary one.
In this situation 
our axion essentially represents a Weinberg-Wilczek axion 
of the mirror world with quite a large mass, 
while it couples the ordinary particles 
like an invisible axion.  
The experimental and astrophysical limits are discussed  
and an allowed parameter window is found 
with the Peccei-Quinn scale $f_a\sim 10^4-10^5$ GeV 
and the axion mass $m_a\sim 1$ MeV, which can be accessible 
for future experiments.  
We also show that our solution to the strong 
CP-problem is stable against the Planck scale induced 
effects.  

\end{abstract}

\end{titlepage}

\section{Introduction}

The strong CP problem remains one of the puzzling points of 
the modern particle physics (for a review, see e.g. ref. \cite{kim}). 
It is related to the P- and CP- violating term $\theta G\tilde{G}$ 
in the QCD Lagrangian which is contributed also by the complex phases 
in the quark mass matrices $M_{U,D}$, 
so that its effective value becomes  
$\overline{\theta}=\theta+ \arg(\det M_U \det M_D)$. 
The CP-violating phenomena in weak interactions indicate 
that the CP-violating phase in the 
Cabibbo-Kobayashi-Maskawa (CKM) matrix is large,  
and hence the quark mass matrices should contain large 
complex phases. Therefore, one would generally expect that 
$\overline{\theta}$ is order 1. 
However, the experimental bound on the neutron electric dipole 
moment yields $\overline{\theta} < 10^{-10}$.  
The strong CP problem simply questions why the effective 
$\overline{\theta}$-term is so small, 
or in other words, what is the origin of such a strong 
fine tuning between the initial value $\theta$ and 
the phases of the quark mass matrices? 

The most appealing solution to the problem is provided by the
Peccei-Quinn (PQ) mechanism \cite{PQ}, based on the concept 
of spontaneously broken global axial symmetry $U(1)_{\rm PQ}$.  
In the context of this mechanism $\overline{\theta}$  
essentially becomes a dynamical field, 
$\overline{\theta}=a/f_a$, with an effective 
potential induced by the non-perturbative QCD effects  
which fixes vacuum expectation value (VEV) at zero.   
Here $a$ stands for the pseudo-Goldstone mode of the 
spontaneously broken PQ symmetry, an axion,  
and $f_a$ is 
a VEV of scalar (or a VEV combination of several scalars) 
responsible for the $U(1)_{\rm PQ}$ symmetry breaking. 
This scale is model dependent. 
In particular, 
in the original Weinberg-Wilczek (WW) model \cite{WW}  
$f_a$ is order electroweak scale,  
but in the invisible axion models  
\cite{DFSZ,KSVZ,archion} it should be much larger. 

In the most general context, the axion couplings to fermions 
and photons are described by the following Lagrangian terms: 
\begin{equation}  \label{a-couplings}
\mathcal{L}_a = ig_{ak}a \bar\psi_k\gamma_5\psi_k + 
{\frac{g_{a\gamma }}{4}} aF_{\mu\nu}\tilde F^{\mu\nu} + ....
\end{equation}
where the coupling constants are essentially determined by the 
PQ scale $f_a$.
Typically, up to model dependent order one coefficients, 
we have $g_{ak}\sim f_a^{-1}m_k$, where $m_k$ is the fermion mass
(e.g. $m_k=m_e,m_N,...$ for electron, nucleons, etc.), 
and $g_{a\gamma}\sim (\alpha /2\pi) f_a^{-1}$, with $\alpha$ 
being the fine structure constant. 

As for the axion mass, it also depends on the QCD scale 
$\Lambda$ where strong interactions become nonperturbative. 
In the theory with no light quarks (with masses below $\Lambda$), 
the axion mass would emerge due to instanton-induced potential 
which can be computed in the dilute gas approximation, 
$P(a) \approx -K\cos(a/f_a)$ with $K\sim \Lambda^4$,
and so we would have $m^2_a\sim \Lambda^4/f_a^2$. 
However, in reality   
the axion state $a$ gets small mixing with the
pseudo-Goldstone bosons $\pi,\eta$, etc., related to the 
chiral symmetry breaking by the light quark 
condensates $V=\langle \bar{q}q\rangle \sim \Lambda^3$. 
In order to find the true mass eigenvalue
corresponding to the axion, the total mass matrix 
should be diagonalized. In doing so, one finds a general 
expression \cite{Choi}: 
\begin{equation}  \label{B}
m_a^2 = \frac{N^2}{f_a^2} \frac{VK}{V + K\mathrm{Tr}M^{-1} }
\end{equation}
where $N$ stands for the color anomaly of $U(1)_{\rm PQ}$ 
current,\footnote{
The PQ charges are normalized so that each of 
the standard fermion families contributes as $N=1$.   
Hence, in the WW model we have $N=N_g$, 
where $N_g(=3)$ is the number of fermion families.  
The same holds in the Dine-Fischler-Srednicki-Zhitnitskii 
(DFSZ) model \cite{DFSZ}. 
Other models of the invisible axion, e.g. 
the hadronic axion \cite{KSVZ} or archion \cite{archion}, 
generally contain some exotic fermions and so 
$N\neq N_g$. } 
and $M= \mathrm{diag}(m_u,m_d,...)$ is a mass matrix 
of light quarks, with $m_q < \Lambda$.  
Thus, the axion mass is given as 
$m_a^2 \sim m_q\Lambda^3/f^2_a$. 
More precisely,
neglecting the $s$-quark contribution and using the relation 
$(m_u + m_d)\langle \bar{q}q\rangle = m_\pi^2 f_\pi^2$, 
from (\ref{B}) one directly arrives to
the more familiar formula: 
\begin{equation}  \label{a-m}
m_a = \frac{N}{f_a} \left(\frac{m_u m_dV}{m_u+m_d}\right)^{1/2} = 
\frac{Nz^{1/2}}{1+z}\cdot \frac{f_\pi m_\pi}{f_a} 
\approx \left(\frac{10^6~\mathrm{GeV}}{f_a/N}\right)  
\times 6.2 ~\mathrm{eV}
\end{equation}
where $z=m_u/m_d\simeq 0.57$, and $m_\pi,f_\pi$ respectively 
are the pion mass and decay constant. 

In the  WW model \cite{WW} the PQ symmetry is broken  
by two Higgs doublets $H_{1,2}$ with the VEVs $v_{1,2}$,
namely $f_a = (v/2)\sin 2\beta$, where 
$v=(v_1^2 + v_2^2)^{1/2}\simeq 247$ GeV is 
the electroweak scale and $\tan\beta=v_2/v_1$. 
Therefore, the WW axion is quite heavy, and its mass can 
vary from few hundred keV to several MeVs:
\begin{equation}  \label{WW-mass}
m_a^{\mathrm{WW}} = 
\frac{2N}{v\sin 2\beta} \left(\frac{m_u V}{1+z}\right)^{1/2} 
\approx \frac{150 ~\mathrm{keV} }{\sin 2\beta}  ,
\end{equation}
However, its couplings to fermions and photon are too strong
and by this reason the WW model is completely ruled out for any 
values of the parameter $\beta$ by a variety of the terrestrial 
experiments as are the search of the decay $K^+\to \pi^+ a$, 
$J/\psi$ and $\Upsilon$ decays into $a+\gamma$, 
nuclear deexcitations via axion emission, the reactor and beam
dump experiments, etc. \cite{PDG}.

For a generic invisible axion, these experimental data imply a 
lower bound $f_a > 10^4$ GeV or so, which leads to the upper 
limit on $m_a$ of about 1 keV. 
However, for an axion being so light, the stringent bounds 
emerge from astrophysics and cosmology. 
In particular, for the DFSZ model the combined limits 
from the stellar evolution and the supernova neutrino signal 
exclude all scales $f_a$ up to $10^{10}$ GeV 
\cite{kim,raffelt}, which in turn implies 
$m_a < 10^{-4}$ eV.\footnote{
In the case of the hadronic axion \cite{KSVZ} or archion 
\cite{archion}, a small window around $f_a \sim 10^6$ GeV 
and so the axion mass of few eV can be also permitted.
}
On the other hand, the cosmological limits related to the 
primordial oscillations of the axion field or to the non-thermal 
axion production by cosmic strings, demand the upper bound 
$f_a < 10^{10-11}$ GeV \cite{kim,raffelt}. 
Thus, not much parameter space remains available. 

In addition, the invisible axion models have a naturality 
problem related to the Planck scale induced terms. 
Many arguments suggest that 
the non-perturbative quantum gravity effects do not respect 
the global symmetries \cite{kamion} and thus they can induce 
the higher order effective operators cutoff by $M_{P}$ 
which explicitly violate the PQ symmetry.
For large $f_a$ these explicit terms can be very dangerous 
for the stability of the PQ mechanism.  
For example, in the DFSZ model where $U(1)_{\rm PQ}$ 
is broken by a gauge singlet scalar $S$,  
the operator $\frac{S^5}{M_P} + \mathrm{h.c.}$  
would induce $\bar{\theta} > 10^{-10}$ unless 
$f_a < 10$ GeV \cite{kamion}. 
Therefore, the Planck scale induced effects  
leave no room for the invisible axion models.
On the other hand, the WW model does not have this problem  
and neither it is sensitive to astrophysical constraints,  
but it is excluded by the laboratory limits because 
of too strong couplings with the matter.\footnote{
There are also "axionless" models where the solution to 
the strong CP-problem is due to spontaneously broken 
P- or CP-invariance \cite{Parity}.
These scenarios, however, have many specifics and do 
not work in the general context. 
The Planck scale induced effects 
for the axionless models have been studied in ref. \cite{BMS}.} 

In the present paper we suggest a new model for the axion. 
We assume that there exists a parallel sector of "mirror" 
particles \cite{mirror,PLB98,BM,BDM} 
which is completely analogous to the ordinary particle sector, 
i.e. it has the same gauge group and all coupling constants 
are equal in two sectors. In other words, the whole Lagrangian 
is invariant with respect a discrete mirror parity (M-parity)
under the exchange of two sectors. 
We further assume that the ordinary and mirror worlds have 
the common PQ symmetry realized \textit{\`a la} Weinberg-Wilczek, 
with the $U(1)_{\rm PQ}$ charges carried by the 
ordinary Higgs doublets $H_{1,2}$ and their
mirror partners $H^{\prime}_{1,2}$.\footnote{
The possibility to use mirror symmetry for solving the 
strong CP problem, in different context, 
was first considered by Rubakov \cite{Rub}. 
We will briefly comment on this model later. 
In principle, two sectors could have also other common symmetries, 
e.g. the flavour symmetry \cite{PLB98}.}

If M-parity is an exact symmetry, then the particle physics 
should be exactly the same in two worlds: in particular, 
the initial $\theta$-terms are equal, $\theta= \theta^{\prime}$, 
the quark mass matrices are identical, $M_{U,D}=M^{\prime}_{U,D}$, 
the QCD scales coincide, $\Lambda =\Lambda^{\prime}$, 
and the axion couples both QCD sectors in the same way: 
$f_a^{-1}a(G\tilde{G} + G^\prime\tilde{G}^\prime)$,  
so that their non-perturbative dynamics should produce 
the same contributions to the axion effective potential. 
Clearly, in this case the strong CP-problem is 
simultaneously solved in both worlds   
-- the axion VEV cancels the $\theta$-terms 
both in the mirror and ordinary sectors.  
In such a realization, however, 
$f_a$ remains order $100$ GeV and thus it is excluded
by the same phenomenological grounds as the original WW model.

As it was suggested in ref. \cite{BM}, the M-parity can 
be spontaneously broken and the electroweak symmetry breaking 
scale $v^{\prime}$ in the mirror sector can be larger than 
the ordinary one $v$. 
This would lead to somewhat different particle physics in 
the mirror sector and it is not a priori clear that the strong 
CP-problem still can be simultaneously fixed in both sectors. 
However, we show below that this is just the case! The simple
reason is that the softly broken M-parity leads to the 
VEV difference between the ordinary and mirror Higgs doublets, 
but the structure of the Yukawa couplings remains 
the same in two sectors. 

As far as $U(1)_{\rm PQ}$ is a common symmetry for two sectors, 
now the PQ scale is determined by the larger VEV 
$v^{\prime}$, i.e. $f_a\sim (v^{\prime}/2)\sin 2\beta^\prime$. 
The axion state $a$ dominanly comes from the mirror Higgs 
doublets $H^{\prime}_{1,2}$, up to small ($\sim v/v^\prime$) 
admixtures from the ordinary Higgses $H_{1,2}$. 
Hence, it is a WW-like axion with respect to mirror sector,
while it couples the ordinary matter 
as an invisible DFSZ-like axion. 
Therefore, the experimental limits on the axion search  
require $f_a> 10^4$ GeV or so. 

However, there are good news concernin the axion mass 
-- for a given scale $f_a$ it appears to be much larger 
than the mass of the ordinary DFSZ axion (\ref{a-m}). 
The reason can be briefly explained as follows. 
Due to the larger weak scale, $v^\prime \gg v$,  
the mirror quark masses are scaled up 
by a factor $v^\prime/v$ with respect to the ordinary quarks,  
$m^{\prime}_{u,d,...} \sim (v^\prime/v) m_{u,d,...}$. 
By this reason, also the mirror QCD scale becomes somewhat 
larger than the ordinary one: $\Lambda^{\prime}> \Lambda$ 
\cite{BM,BDM}. Therefore, the axion potential dominantly 
emerges from the mirror non-perturbative dynamics 
and we have  
$m_a^2 \sim m^\prime_q \Lambda^{\prime 3}/f_a^2$  or 
$m_a^2 \sim \Lambda^{\prime 4}/f_a^2$  
(depending on whether the mass of lightest mirror quark 
$m^\prime_q$ is smaller or larger than $\Lambda^\prime$).  
Therefore, our axion can be much heavier than 
the ordinary invisible axion with 
$m_a^2 \sim m_q \Lambda^{3}/f_a^2$.   
In particular, if with increasing $v^\prime$ also 
$\Lambda^{\prime}$ grows fastly enough, 
our axion can maintain the 
weight category of the WW axion (\ref{WW-mass})  
and its mass can remain in the MeV range even for  
$f_a > 10^4$ GeV.

The paper is organized as follows. 
In sect. 2 we present our model and show how it solves the 
strong CP-problem. We also compute the axion mass and its
couplings to the fermions and photons. In sect. 3 we confront  
our model with the experimental and astrophysical bounds 
and demonstrate that there is an allowed parameter space which 
can be of interest for the future experimental search. 
The stability of our model against the Planck scale
induced effects is discussed in sect. 4. 
Finally, in sect. 5, we summarize our results.

\section{Mirror World and the Peccei-Quinn symmetry}

It was suggested a long time ago that there can exist 
a parallel (mirror) gauge sector of particles and interactions 
which is the exact copy of the visible one and it 
communicates the latter through the gravity and perhaps 
also via some other interactions. 
Various phenomenological and astrophysical implications 
of this idea have been studied in refs. 
\cite{mirror,PLB98,BM,BDM}.

In particular, one can consider a model based on the gauge 
symmetry $G\times G^{\prime}$ where 
$G=SU(3)\times SU(2)\times U(1)$ stands for the standard
model of the ordinary particles: the quark and lepton fields 
$\psi_i =q_i,~l_i, ~u^c_i, ~d^c_i, ~e^c_i$ 
($i$ is a family index) and two Higgs doublets $H_{1,2}$, 
while 
$G^{\prime}=SU(3)^\prime\times SU(2)^\prime
\times U(1)^{\prime}$ is its 
mirror gauge counterpart with the analogous particle content: 
the fermions 
$\psi^{\prime}_i= q^{\prime}_i,~l^{\prime}_i, ~u^{c\prime}_i,
~d^{c\prime}_i, ~e^{c\prime}_i$ and the Higgses $H^{\prime}_{1,2}$. 
From now on all fields and quantities of the mirror sector have 
an apex to distinguish from the ones belonging to the ordinary one. 
All fermion fields 
$\psi,\psi^{\prime}$ are taken in a left-chiral basis.

Let us assume that the theory is invariant under 
the mirror parity M: $G\leftrightarrow G^{\prime}$, 
which interchanges all corresponding representations of 
$G$ and $G^{\prime}$. Therefore, the two sectors are 
described by identical Lagrangians 
and all coupling constants (gauge, Yukawa, Higgs) 
have the same pattern in both of them. 
In particular, for the Yukawa couplings 
\begin{eqnarray}  \label{SM-Yuk}
&& \mathcal{L}_{\mathrm{Yuk}} = 
G_U^{ij} u^c_i q_j H_2 + G_D^{ij} d^c_i q_j H_1 + 
G_E^{ij} e^c_i l_j H_1 + {\rm h.c.}, \nonumber \\
&& \mathcal{L}^{\prime}_{\mathrm{Yuk}} = 
G_U^{\prime ij} u^{c\prime}_i q^{\prime}_j H^{\prime}_2 + 
G_D^{\prime ij}d^{c\prime}_i q^{\prime}_j H^{\prime}_1 + 
G_E^{\prime ij}e^{c\prime}_i l^{\prime}_j H^{\prime}_1
+{\rm h.c.} 
\end{eqnarray}
we have $G_{U,D,E}^{ij} = G^{\prime ij}_{U,D,E}$.  
In addition, also the initial $\theta$-terms are equal, 
$\theta = \theta^\prime$. 

We further assume that two sectors have a common Peccei-Quinn symmetry 
$U(1)_{\rm PQ}$ under which fermions $\psi_i,\psi^{\prime}_i$ 
change their phases by a factor $\exp(-i\omega/2)$ and 
the Higgses $H_{1,2}, H^{\prime}_{1,2}$ change
phases as $\exp(i\omega)$. 
Then the renormalizable Higgs Potential has a general form 
$\mathcal{V}_{\mathrm{tot}} = 
\mathcal{V} + \mathcal{V}^{\prime}+ \mathcal{V}_{\mathrm{mix}}$, 
where for the ordinary Higgses we have: 
\begin{equation}  \label{pot1}
\mathcal{V} = -\mu_1^2 H_1^\dagger H_1 -\mu_2^2 H_2^\dagger H_2 +\lambda_1
(H_1^\dagger H_1)^2 + \lambda_2 (H_2^\dagger H_2)^2 + \lambda (H_1
H_2)^\dagger (H_1 H_2),
\end{equation}
the mirror Higgs potential $\mathcal{V}^{\prime}$ has 
exactly the same pattern by M-parity: 
$H_{1,2} \to H^{\prime}_{1,2}$, and the mixing terms are: 
\begin{equation}  \label{pot-mix}
\mathcal{V}_{\mathrm{mix}} = -\kappa (H_1 H_2)^\dagger 
(H^{\prime}_1 H^{\prime}_2) ~ + ~ \mathrm{h.c.}
\end{equation}
where the coupling constant $\kappa$ should be real due to 
M-parity.\footnote{Notice, that 
in the limit $\kappa =0$ there emerge two separate axial 
global symmetries, $U(1)_A$ for ordinary sector under 
which $\psi_i \to \exp(-i\omega/2)\psi_i$ and 
$H_{1,2} \to \exp(i\omega) H_{1,2}$,
 and $U(1)^{\prime}_A$ for mirror sector: 
$\psi^{\prime}_i \to \exp(-i\omega^{\prime}/2)\psi^{\prime}_i$ 
and $H^{\prime}_{1,2} \to \exp(i\omega^{\prime}) H^{\prime}_{1,2}$. 
Therefore, the term $\mathcal{V}_{\mathrm{mix}}$ demands that
$\omega^{\prime}=\omega$ and thus it reduces $U(1)_A\times
U(1)^{\prime}_A$ to its diagonal subgroup $U(1)_{\rm PQ}$.}

The mirror parity can be spontaneously broken so that 
the weak interaction scales 
are different in two sectors \cite{BM,BDM}. 
The easiest way is to introduce a real singlet scalar 
$\eta$ which is odd under M-parity: $\eta \to -\eta$ \cite{Parida}. 
Then the following renormalizable terms can be added 
to the Higgs potential: 
\begin{equation}  \label{pot-eta}
\Delta\mathcal{V} = h(\eta^2 - \mu^2)^2 + 
\sigma_k\eta^2(H_k^\dagger H_k +H^{\prime\dagger}_kH^{\prime}_k) 
+\rho_k\mu \eta 
(H_k^\dagger H_k - H^{\prime \dagger}_k H^{\prime}_k),  
~~~~k=1,2
\end{equation}
A non-zero VEV $\langle\eta\rangle=\mu$ spontaneously 
breaks the M-parity and induces the different 
effective mass$^2$ terms for the ordinary and mirror Higgses:
$-\mu^{2}_{k(\mathrm{eff})} = -\mu_{k}^2 + \sigma_{k}\mu^2  
+ \rho_{k}\mu^2 $ and 
$-\mu^{\prime 2}_{k(\mathrm{eff})} 
= -\mu^2_k+ \sigma_{k} \mu^2 -\rho_{k}\mu^2$. 
Therefore, the mirror VEVs $v^{\prime}_{1,2}$ 
are different from the ordinary ones $v_{1,2}$, 
and in general also the VEV ratios 
$v_2/v_1 =\tan\beta \equiv x$ and 
$v^{\prime}_2/v^{\prime}_1 =\tan\beta^{\prime}\equiv x^{\prime}$ 
must be different in two sectors.
In particular, we will be interested in a situation when  
$v^{\prime}_{1,2}\gg v_{1,2}$, i.e. the weak interaction scale 
$v^{\prime}= (v^{\prime 2}_1+v^{\prime 2}_2)^{1/2}$ in the
mirror sector is much larger than the ordinary weak scale 
$v = (v^2_1+v^2_2)^{1/2}=247$ GeV,  
$v^\prime/v\equiv \zeta \gg 1$ \cite{BM,BDM}.
Clearly, the large hierarchy between $v^{\prime}$ and $v$ 
requires the fine tuning of the parameters in the Higgs 
potential. We do not discuss here this
question and refer the reader e.g. to ref. \cite{BDM}.  

As far as the ordinary and mirror Yukawa constants 
in (\ref{SM-Yuk}) are the same, the quark and lepton mass 
matrices essentially have the same form in both sectors: 
$M_U = G_U \langle H_2 \rangle$, 
$M^\prime_U = G_U \langle H^\prime_2 \rangle$, 
$M_D = G_D \langle H_1 \rangle$, 
$M^\prime_D = G_D \langle H^\prime_1 \rangle$, etc. 
Thus, the fermion mass and mixing structure in the 
mirror sector should be the same as in the ordinary one, 
with the mirror up quark masses scaled as 
$m^{\prime}_{u,c,t}= \zeta_2 m_{u,c,t}$  
and the down quark and charged lepton masses scaled as 
$m^{\prime}_{d,s,b}= \zeta_1 m_{d,s,b}$, 
$m'_{e,\mu,\tau}= \zeta_1 m_{e,\mu,\tau}$, where 
$\zeta_2 = v^\prime_2/v_2=\zeta(\sin\beta^\prime/\sin\beta)$ 
and $\zeta_1 = v^\prime_1/v_1=\zeta(\cos\beta^\prime/\cos\beta)$.

At very high energies, $\mu \gg v^{\prime}$,
the strong coupling constants $\alpha_s(\mu)$ and 
$\alpha^{\prime}_s(\mu)$ should be equal due to M-parity.  
Under the renormalization group (RG) evolution 
they both evolve down in parallel ways until
the energy reaches the value of mirror-top mass 
$m^{\prime}_t \simeq \zeta_2 m_t$. 
Below it $\alpha^{\prime}_s$ will have a different slope than 
$\alpha_s$, and this slope will change every time below the 
mirror quark thresholds $m^{\prime}_b \sim \zeta_1 m_b$, etc., 
In the evolution of $\alpha_s$ these thresholds occur at lower  
scales, $\mu = m_t,m_b$, etc. 
Then it is very easy to determine the scale $\Lambda^{\prime}$ 
at which $\alpha^{\prime}_s$ becomes large, 
once we know that for the ordinary QCD this happens at 
$\Lambda\simeq 200$ MeV. In other words, $\Lambda'$ 
becomes a function of $\zeta_{1,2}$, and   
for $v^{\prime}\gg v$ one could obtain a significant 
difference between the QCD scales: 
$\Lambda^{\prime}>\Lambda$. 

Let us consider for simplicity the case $x=x'$, 
which yields $\zeta_1=\zeta_2=v'/v$.  
Then for the values of $v^{\prime}$ up to $10^4$ GeV 
or so one obtains an approximate scaling low: 
\begin{equation}  
\label{xi}
\frac{\Lambda^\prime}{\Lambda} 
\simeq \left( \frac{v^\prime}{v} \right)^{\frac{b_2-b_6}{b_2}} 
\simeq \left(\frac{v^\prime}{v}\right)^{0.28}
\end{equation}
where $b_n= 11-\frac23 n$ are the RG coefficients for the case 
of $n$ light quarks. For $v^{\prime}> 10^4$ GeV the mirror light 
quarks masses become larger than $\Lambda^{\prime}$, 
since they grow faster with $v^{\prime}$: 
$m^{\prime}_{u,d}/m_{u,d}=v'/v$, and the approximation 
(\ref{xi}) is not valid -- for $\mu < m^\prime_{u,d}$ 
the mirror QCD becomes the pure gluodynamics and hence
$\Lambda^{\prime}/\Lambda$ starts to
increase faster with $v^{\prime}$ (see Fig. 1).

\begin{figure}[t]
\par
\begin{center}
\includegraphics[width=10cm,angle=0]{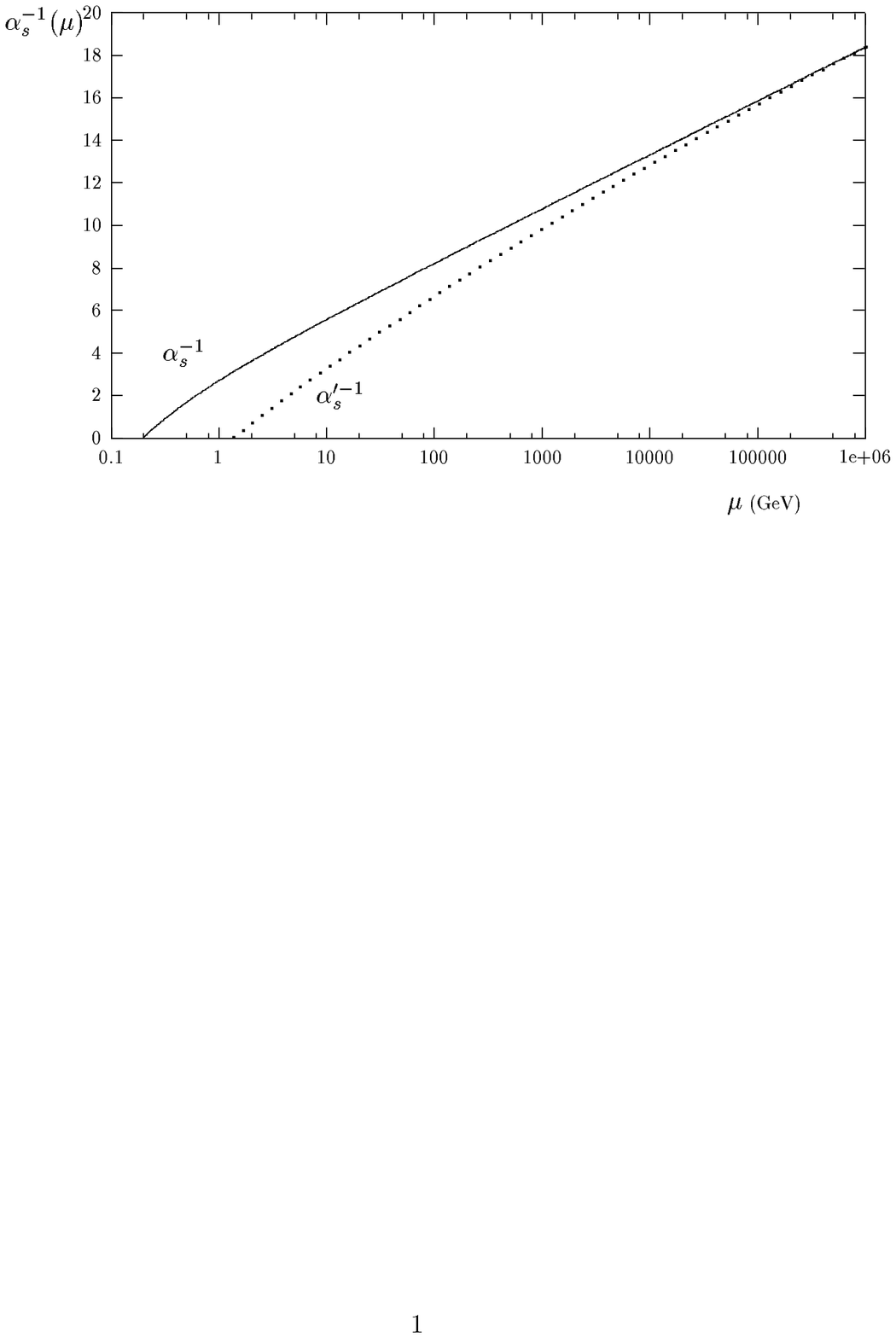}
\end{center}
\par
\vspace{-5mm}
\caption{{\protect\small Renormalization group evolution of the strong
coupling constants $\alpha_s$ and $\alpha^{\prime}_s$ from higher to lower
energies, for $v^{\prime}= 10^6$ GeV and $x=x^{\prime}$. }}
\end{figure}

Let us discuss now the axion physics in our model. 
Since we are interested only in Goldstone modes, let us express 
the neutral components of the ordinary and mirror Higgs
fields as: 
\begin{equation}  \label{Neutr}
H^0_{k}=\frac{v_k}{\sqrt{2}} \exp(i\phi_k/v_k), ~~~~~ 
H^{\prime 0}_{k}=\frac{v^{\prime}_k}{\sqrt{2}} 
\exp(i\phi^{\prime}_k/v^{\prime}_k); ~~~~~ k=1,2
\end{equation}
and identify the combination of the phases $\phi_{1,2}$ and 
$\phi^{\prime}_{1,2}$ which corresponds to the axion field. 
We immediately 
observe that the states  
$\phi_Z = v^{-1}(v_2 \phi_2 - v_1 \phi_1)$ and 
$\phi^{\prime}_Z = v^{\prime -1}(v^{\prime}_2 \phi^{\prime}_2 - 
v^{\prime}_1\phi^{\prime}_1)$ are the true Goldstone modes 
eaten up respectively by the standard and mirror gauge bosons
$Z$ and $Z^{\prime}$  which become massive
particles by the Higgs mechanism.

In the limit $\kappa=0$, two remaining combinations 
$\phi = v^{-1}(v_1 \phi_2 + v_2 \phi_1)$ and 
$\phi^{\prime}= v^{\prime -1}(v^{\prime}_1 \phi^{\prime}_2 +
v^{\prime}_2 \phi^{\prime}_1)$ 
would represent the Goldstone modes of larger
global symmetry $U(1)_A\times U(1)^{\prime}_A$, 
respectively with the decay constants 
$f= v_1v_2/v = (v/2)\sin 2\beta$ and 
$f^{\prime}=v^{\prime}_1v^{\prime}_2/v^{\prime}=
(v^{\prime}/2)\sin 2\beta^{\prime}$. 
Thus, we would have two axion states separately 
for the ordinary and mirror sectors, and
their masses would emerge respectively due to quantum 
anomalies 
of the $U(1)_A$ and $U(1)^{\prime}_A$ currents related to 
ordinary and mirror QCD.
In particular, $\phi$ would represent the familiar WW axion with 
order $\sim 1$ MeV mass given by eq. (\ref{WW-mass}).

However, the mixing term (\ref{pot-mix}) reduces the larger 
global symmetry $U(1)_A\times U(1)^{\prime}_A$ to a common PQ 
symmetry $U(1)_{\rm PQ}$.
Substituting (\ref{Neutr}) in $\mathcal{V}_{\mathrm{mix}}$, 
we obtain: 
\begin{equation}  \label{A-pot}
\mathcal{V}_{\mathrm{mix}} = 
-\frac{\kappa}{2} v_1 v_2 v^{\prime}_1v^{\prime}_2 
\cos \left(\frac{\phi}{f} - \frac{\phi^{\prime}}{f^{\prime}}\right) 
= \frac{\kappa v v^{\prime}}{4ff^{\prime}}
(f^{\prime}\phi - f\phi^{\prime})^2 + \dots
\end{equation}
Thus, the combination 
$A = f_a^{-1} (f^{\prime}\phi -f\phi^{\prime})$, 
where $f_a =(f^2 + f^{\prime 2})^{1/2}$, 
receives a mass$^2$ term: 
\begin{equation}  \label{A-mass}
M_A^2 = \frac{\kappa f_a^2}{2} \left(x+ \frac{1}{x}\right) 
\left(x^{\prime}+ \frac{1}{x^{\prime}}\right) = 
\frac{2\kappa f_a^2}{\sin 2\beta \sin 2\beta^{\prime} } 
\end{equation}
which is \textit{positive} for $\kappa > 0$. 
The present experimental limits on the Higgs search imply 
the lower bound on $M_A$ about 100 GeV \cite{PDG}. 
As for the other combination 
$a= f_a^{-1} (f \phi + f^{\prime}\phi^{\prime})$, it remains 
as a Goldstone mode of the $U(1)_{\rm PQ}$ symmetry to be 
identified with the axion field.

Let us consider now the effective $\theta$-terms in two sectors. 
Recalling that 
$\theta=\theta^{\prime}$ and $G_{U,D} = G_{U,D}^{\prime}$  
due to M-parity,   
we obtain:\footnote{
In the following, 
we maintain parametrically the number of fermion families,
$N=N_g$, in general formulas, but for the numerical calculations 
we always take $N=3$.  
}
\begin{eqnarray}  \label{Thetas}
&&
\bar\theta = \theta + \arg(\det M_U \det M_D) =   
\theta + \arg(\det G_U \det G_D) + N \langle\phi/f\rangle   
\nonumber \\
&&
\bar\theta^{\prime}= 
\theta^{\prime}+ \arg(\det M^{\prime}_U \det M^{\prime}_D) = 
\theta + \arg(\det G_U \det G_D) + N 
\langle \phi^{\prime}/f^{\prime}\rangle 
\end{eqnarray}
where the last terms describe contributions of the phases 
in (\ref{Neutr}), and they can be immediately rewritten as 
$\phi^{\prime}/f^{\prime}=f_a^{-1} [a - (f/f^{\prime})A]$ and 
$\phi/f = f_a^{-1} [a + (f^{\prime}/f)A]$. 
We see that the axion field $a$ contributes both eqs. 
(\ref{Thetas}) in the same way, whereas the heavy state 
$A$ is irrelevant since it has a vanishing VEV. 
Therefore, $\bar{\theta}$ and $\bar{\theta}^{\prime}$ 
will be simoultaneously cancelled by the axion VEV 
$f_a^{-1} \langle a\rangle$ and  
so in our model the strong CP-problem is
solved in both sectors.\footnote{
Actually $\bar{\theta}$ can get radiative corrections due to 
the fact that the electroweak scales are different in two 
sectors. However, it is well-known that these corrections are 
negligibly small \cite{Yulik}.}

In addition, for the case $f^{\prime}\gg f$ we obtain 
a quite interesting situation.
The axion state $a$ dominantly emerges from the phases of 
the mirror Higgses $H^{\prime}_{1,2}$ while it contains only 
a small admixture from the ordinary Higgses $H_{1,2}$: 
$a \approx \phi^\prime - (f/f^\prime)\phi$. 
Therefore, it appears to be a WW-like axion with respect 
to the mirror sector, 
while from the view of the ordinary sector it is just a
DFSZ-like invisible axion with a decay constant 
$f_a \approx f^{\prime}=
(v^{\prime}/2)\sin 2\beta^{\prime}$.\footnote{
This is very easy to understand from the following observation. 
The combination $H^{\prime}_1H^{\prime}_2\equiv S$ 
in (\ref{pot-mix}) is a standard model 
singlet carrying the double PQ charge,  
and it can be considered as a composite operator 
which communicates the standard Higgses $H_{1,2}$ via the term 
$\mathcal{V}_{\mathrm{mix}} =
(H_1H_2)^\dagger S + \mathrm{h.c.}$, just as in the DFSZ model.}

However, there is a dramatic difference from the DFSZ case. 
Since the mass of our axion receives
contributions from both ordinary and mirror QCD, 
in a full analogy to eq. (\ref{B}) one can write: 
\begin{equation}  \label{B'}
m^2_a = \frac{N^2}{f^2_a} 
\left(\frac{VK}{V + K\mathrm{Tr}M^{-1} } + 
\frac{V^{\prime}K^{\prime}}{V^{\prime}+ 
K^{\prime}\mathrm{Tr}M^{\prime -1} } \right)
\end{equation}
where $M = \mathrm{diag}(m^{\prime}_u, m^{\prime}_d)$ 
is the mass matrix of the mirror light quarks, 
and the values $K^{\prime}\sim \Lambda^{\prime 4}$ and  
$V^{\prime}\sim \Lambda^{\prime 3}$ characterize the 
mirror gluon and quark condensates, with $\Lambda^{\prime}$ 
being the QCD scale in the mirror sector. 
The first term in this expression gives an ordinary 
{\it small} contribution (\ref{a-m}) and it is negligible  
in comparison to the second term which can give {\it much larger} 
contribution as far as $m'_{u,d} \gg m_{u,d}$ and 
$\Lambda' > \Lambda$. Certainly, the exact form of the latter  
term depends on how many light quarks 
(with masses less than $\Lambda^{\prime}$) are 
contained in mirror sector. 
In particular, for the cases of two ($u^{\prime}$
and $d^{\prime}$), one ($u^{\prime}$), or zero light quarks, 
the second term in brackets of eq. (\ref{B'}) becomes respectively  
$(1+z^\prime)^{-1}m^{\prime}_uV^{\prime}$,
$m^{\prime}_u V^{\prime}$ or $K^{\prime}$, where 
$z^\prime=m^\prime_u/m^\prime_d= (\zeta_2/\zeta_1) z=
(x^\prime/x)z$. 

Let us take for example the case $x=x^{\prime}$, 
and consider the regime when 
at least $u^{\prime}$ is light, 
$m^{\prime}_{u}= (v'/v)m_{u} < \Lambda^{\prime}$  
(recall that this regime holds up to the values 
$v^\prime\sim 10^4$ GeV or so).  
Then for the axion mass holds an approximate scaling low: 
\begin{equation}  \label{case}
m_a \approx \frac{2N}{v^\prime\sin 2\beta^\prime}
\left(\frac{m^\prime_uV^\prime}{1+z^\prime}\right)^{1/2} 
\approx \left(\frac{\Lambda^{\prime}}{\Lambda}\right)^{3/2} 
\left(\frac{v}{v^{\prime}}\right)^{1/2} m_a^{\mathrm{WW}}. 
\end{equation}
with respect the ordinary WW axion mass (\ref{WW-mass}).    
Using also eq. (\ref{xi}), we thus obtain:  
\begin{equation}  \label{m-scale}
m_a \simeq \left(\frac{v}{v^{\prime}}\right)^{0.08} 
\times \frac{150 ~\mathrm{keV} }{\sin 2\beta}  ,
\end{equation}
In general case, 
for very large $v^\prime$ and arbitrary $x^\prime$, the 
simple scaling low (\ref{case}) is no more valid, since  
for $x\neq x^{\prime}$ the masses of $u^{\prime}$ 
and $d^{\prime}$ scale differently and in addition, 
with increasing $v'$, only one or none of them  
may remain light, with mass less than $\Lambda^\prime$).  
In this situation one has to use the general formula (\ref{B'}).  
Nevertheless, $m_a$ still decreases very slowly with 
increasing $v^{\prime}$ and for enough large $x,x^\prime$, 
its value can be of the order of MeV.  

The results of numerical computations from 
the general formula (\ref{B'}) 
for some interesting cases are presented in Fig. 2. 
The relevant parameters are taken as $x=\tan\beta$, 
$x^{\prime}=\tan \beta^{\prime}$ in the range $1-100$, 
and $f_a=v^{\prime}/(x^{\prime}+x^{\prime -1})$. 
We use one-loop RG evolution for the strong coupling 
constants with $\Lambda=200$ MeV. 
The following values of the quark masses are used: 
$m_u=4$ MeV, $m_d=7$ MeV, $m_s=150$ MeV (at $\mu=1$
Gev), and $m_c(m_c)=1.3$ GeV, $m_b=4.3$ GeV 
and $m_t=170$ GeV (respectively at $\mu=m_c,m_b,m_t$). 
For the parameters $V$ and $K$ related to the quark and 
gluon condensates we take $V = (250 ~\mathrm{MeV})^3$ and 
$K = (230 ~\mathrm{MeV})^4$, and assume that the similar
parameters in the mirror sector scale as $V^{\prime}/V =
(\Lambda^{\prime}/\Lambda)^3$ and $K^{\prime}/K =
(\Lambda^{\prime}/\Lambda)^4$. We see that for enough 
large values of $x,x'$ the axion mass can be order  
$1$ MeV even for $f_a = 10^4-10^6$ GeV. 

Let us discuss now the interaction terms (\ref{a-couplings}). 
Clearly, our axion couples the ordinary
fermions and photons in exactly the same way 
as the DFSZ axion with the PQ scale $f_a$. 
Hence, for the Yukawa coupling constants to the electron, 
down quark and up quark  we can use the familiar 
expressions \cite{kim}: 
\begin{equation}  \label{Yuk-eud}
g_{ae} = \frac{m_e}{f_a}\sin^2\beta, ~~~~~ 
g_{ad} =\frac{m_d}{f_a}
\left[\sin^2\beta- \frac{N}{1+z}\right], ~~~~~ 
g_{au} = \frac{m_u}{f_a}
\left[\cos^2\beta- \frac{Nz}{1+z}\right], 
\end{equation}
while for the axion-photon coupling we have: 
\begin{equation}  \label{photon}
g_{a\gamma}= \frac{\alpha}{2\pi} \frac{NC}{f_a}, ~~~~ 
C = \frac83 - \frac{6K \mathrm{Tr}(M^{-1}Q^2)} 
{V + K \mathrm{Tr}(M^{-1}) } \simeq \frac{2z}{1+z}
\end{equation}
where the trace is taken over the light quark states 
($u,d$), $Q$ are their electric charges $(+4/3,-1/3)$, 
and $\alpha=1/137$ is the fine structure constant. 
In addition, our axion couples to mirror photons,  
with the constant:  
\begin{equation}  \label{photon-mir}
g^{\prime}_{a\gamma}= \frac{\alpha}{2\pi} \frac{NC^{\prime}}{f_a}, ~~~~
C^{\prime}= \frac83 - \frac{6K^{\prime}\mathrm{Tr}(M^{\prime -1}
Q^{\prime 2})} {V^{\prime}+ K^{\prime}\mathrm{Tr}(M^{\prime -1})}
\end{equation}
where the factor $C^{\prime}$ for the case of two 
($u^\prime,d^\prime$),  one ($u^\prime$) or no 
light quarks respectively takes 
the values $2z^{\prime}/(1+z^{\prime})$, 0 and $8/3$.  

Hence, the axion decay widths  
into the visible and mirror photons respectively are: 
\begin{equation}  \label{width}
\Gamma(a\to \gamma \gamma)= 
\frac{g^2_{a \gamma} m_{a}^3}{64 \pi}, ~~~~
\Gamma(a\to \gamma^{\prime}\gamma^{\prime})= 
\frac{g^{\prime 2}_{a \gamma}m_{a}^3}{64 \pi},
\end{equation}

In addition, if $m_a >2m_e$, the axion can decay 
also into electron-positron pair: 
\begin{equation}  \label{width-e}
\Gamma(a\to e^+e^-)= \frac{g^2_{ae} m_{a}}{8 \pi} 
\sqrt{1 - \frac{4m_e^2}{m_a^2} }
\end{equation}

\begin{figure}[t]
\includegraphics[width=5.5cm,angle=270]{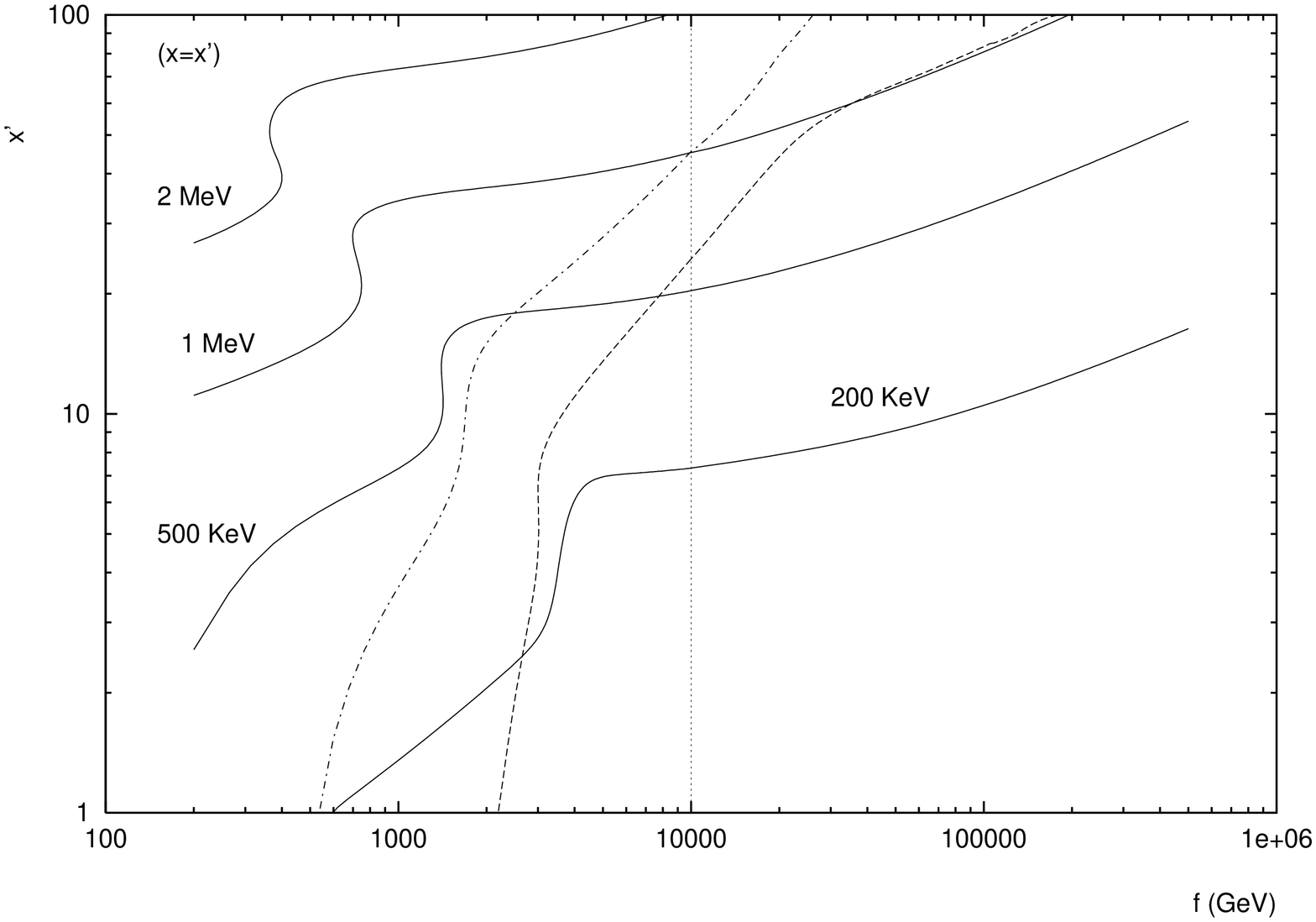} 
\includegraphics[width=5.5cm,angle=270]{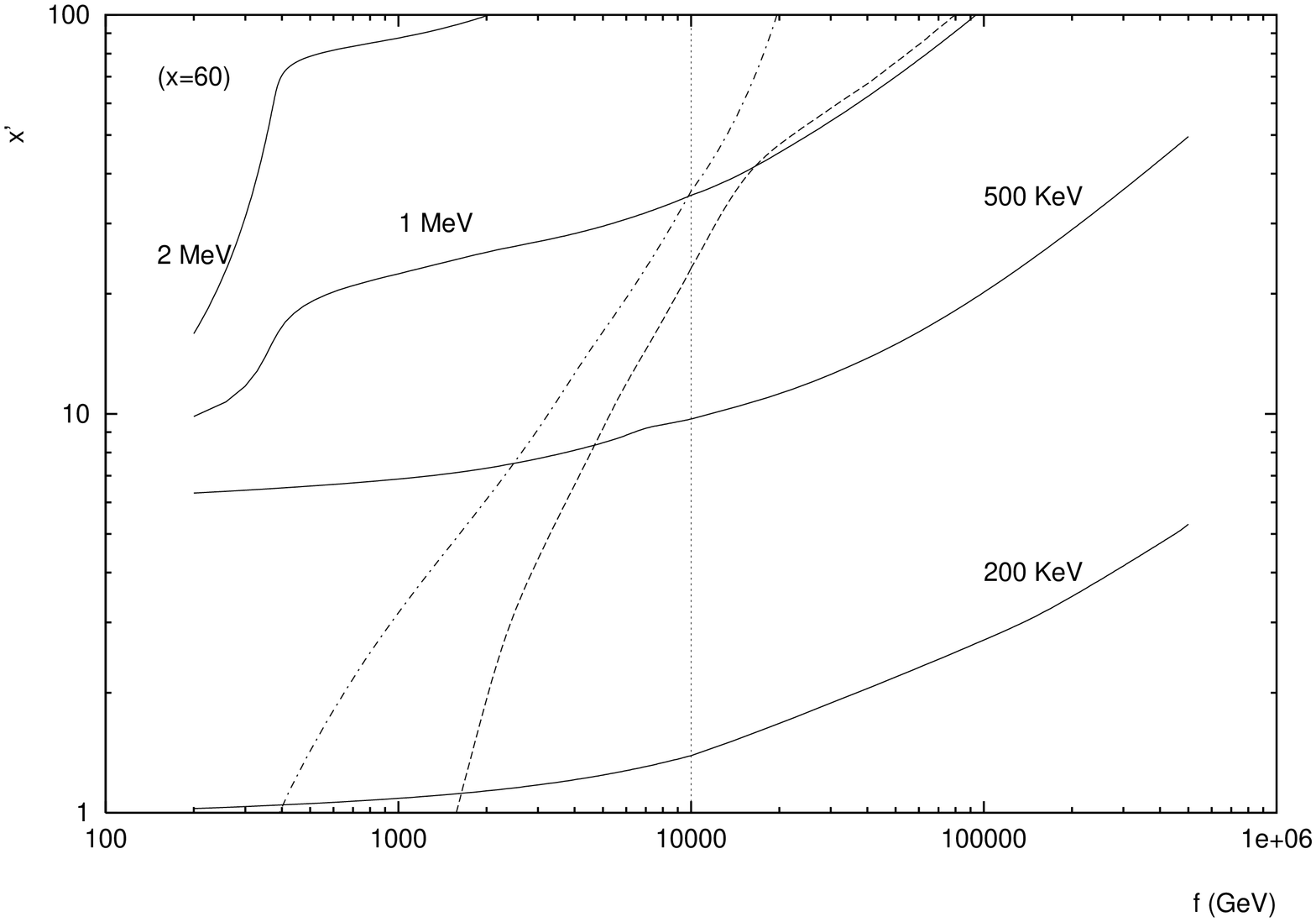} 
\caption{{\protect\small 
Contour plots for the axion mass (solid) as a function 
of $f_a$ and $x^\prime$, for the cases $x=x^\prime$ and 
$x=60$. 
The regions left from the dash-dots curves 
are excluded by the observed neutrino signal 
from SN 1987A while 
the regions right from the dash curves are excluded by 
the SMM bounds on the gamma ray flux from SN 1987A.  
The vertical dott line indicates a conservative bound 
on the PQ scale from the terrestrial experiments. 
}}
\end{figure}

\section{Experimental and astrophysical bounds}

Let us discuss now the constraints which should be 
imposed on the axion mass $m_a$ and 
its coupling constants for fulfilling the
existing experimental and astrophysical bounds. 
Anticipating our results, we remark that the consistent 
parameter region corresponds to large values of $x=\tan\beta$ 
and so $\sin\beta\approx 1$.  
Hence, the axion coupling constants to the electron
is $g_{ae}= 5\times 10^{-10}f_6^{-1}$, and for 
the axion-nucleon coupling constants we have the 
following estimates 
$g_{ap}=-1.6\times 10^{-6}f_6^{-1}$ and 
$g_{an}=0.6\times 10^{-6}f_6^{-1}$ 
(see e.g. in ref. \cite{raffelt}),  
where $f_6 = (f_a/10^6~\mathrm{GeV})$.

The conservative interpretation of the experimental 
constraints from the axion search \cite{PDG} 
imply the lower limit on the axion-nucleon coupling 
constants $g_{aN}< 10^{-4}$ or so, 
which translates as $f_6 > 10^{-2}$. 
For $m_a > 2m_e$, the somewhat stronger limit emerges 
from the reactor search of $a\to e^+e^-$ decay. 
Namely, for $m_a = 1.4-2$ MeV one has $f_6 > 0.1$ \cite{altmann}. 
The latter limit, however, is subject of various theoretical
uncertainties, and in the following we take the most conservative 
lower limit $f_a > 10^4$ GeV. 
The existing limits on the axion decay into two photons 
are also fulfilled in this case \cite{faissner}. 

For the standard DFSZ axion, since it is very very light,   
the severe limits on the scale $f_a$ emerge 
from the stellar evolution. 
In particular, the astrophysical bounds on the 
white dwarf luminosity function, helium burning lifetime 
of the HB stars or helium ignition time in low mass red giants 
imply $g_{ae} < (2.5 - 5)\times 10^{-13}$, 
and hence $f_a > (1-2)\cdot 10^{9}$ GeV \cite{raffelt}. 
However, these limits do not apply to our axion 
which is quite heavy and so    
its production rate in the stellar cores, with typical
temperatures $T$ up to 10 keV, is suppressed by the 
exponential factor $\exp(-m_a/T)$. For $m_a > 300$ keV 
this factor becomes enough small to render our model safe 
as far as the stellar evolution limits are concerned.

However, stringent bounds on $f_a$ emerge due to the 
observed neutrino signal from the SN 1987A 
which indicates that the supernova core cooling rate 
due to axion emission at the typical time $t=1$ s 
from the collapse should not exceed $10^{51}$ erg/s 
or so \cite{raffelt}. 
If the axion-nucleon couplings are enough large, 
$g_{ap},g_{an}> 10^{-7}$, the axions are trapped in the 
collapsing core and are emitted with the thermal spectrum 
from an axiosphere with a radius $R\sim 10$ km, and 
energy luminosity at $t=1$ s can be estimated as
$L_a \simeq f_6^{16/11}\times 3\cdot 10^{50}$ erg/s 
\cite{Tur,Bur}.  
Hence, for $f_6 < 3$ the axion luminosity $L_a$ 
becomes smaller than $10^{51}$ erg/s, and   
the total energy and duration of the SN 1987A neutrino burst 
should not be affected.  
This consideration is certainly applicable for an
axion with mass of order MeV, and so we impose 
the upper limit $f_a < 3\cdot 10^6$ GeV or so.\footnote{
The region $f_a > 10^{10}$ is also allowed by the 
SN 1987A \cite{raffelt}, however it is not of our interest 
in this paper.} 

In addition, the supernova bounds provide strong   
limits on the axion decay rates into the ordinary 
as well as mirror photons.  
For $m_a \sim 1$ MeV these decays are rather fast 
and strongly restrict the allowed parameter space. 

As far as the ordinary matter is transparent 
for the mirror photons, the emission of the latter can lead 
to the unacceptably fast cooling of the supernova core. 
The energy lose rate $L^{\prime}_\gamma$ related to the mirror 
photon emission from the core volume can be estimated as follows. 
For $f_a < 10^7$ GeV the axions are strongly trapped 
inside a core of the radius $R\simeq 10$ km and have 
a thermal distribution. The temperature $T$  
is a function of the radial distance $r$ from the center, 
and it also slowly changes with the time $t$ elapsed from 
the collapse.  
The decay rate for an axion with the energy $E$ into mirror 
photons is $(m_a/E)\Gamma^\prime$, where 
$\Gamma^\prime = \Gamma(a\to \gamma^\prime \gamma^\prime)$ 
is given by eq. (\ref{width}).   
Therefore, we have 
$L^{\prime}_\gamma(t) = 4\pi m_a \Gamma^\prime 
\int_0^R r^2 n_a(r,t) dr$, 
where $n_a = 1.2T^3/\pi^2$ is the axion number density. 
Taking a core temperature $T$ of about $20-30$ MeV, 
one can roughly estimate that  $L^{\prime}_\gamma \simeq 
\Gamma^\prime m_a (4\pi R^3/3)(1.2T^3/\pi^2)\sim 
10^{-2} g^{\prime 2}_{a\gamma} m_a^4$.  
Hence, the condition $L^{\prime}_\gamma < 10^{51}$ erg/s  
implies that the decay width $\Gamma^\prime$,
for $m_a \sim 1$ MeV, should not exceed few s$^{-1}$.  
In other terms, using eq. (\ref{photon-mir}), we obtain 
roughly the bound $m_a^2/f_a < 10^{-7}$ MeV.   

By taking the typical temperature profiles in the supernova 
core at different time moments \cite{raffelt}, one can compute 
the value  $L^{\prime}_\gamma(t) = 
4\pi \Gamma^\prime m_a \int_0^R r^2  n_a(r,t) dr$ 
more precisely.  
In addition, $m_a$ and $g^\prime_{a\gamma}$ can be 
expressed as functions of $x,x^\prime$ and $f_a$ 
according to eqs. (\ref{B'}) and (\ref{photon-mir}).    
In Fig. 2 we show isocontours (dash-dott)   
corresponding to $L^\prime_\gamma = 10^{51}$ erg/s 
at $t=1$ s from the collapse,   
which in fact indicate a lower limit on 
the PQ scale $f_a$ for given $x$ and $x'$.  

Another constraint on our model 
emerges due to the axion decay into ordinary photons.  
For $f_a > 10^{4}$ GeV 
the flux of axions emitted from the collapsing core 
is quite large, and thus if the axions 
would decay into the visible photons in their travel 
to the earth, one would observe the gamma ray burst 
associated with the SN 1987A. 
However, the SMM satellite data \cite{SMM} sets 
the stringent upper limit on the $\gamma$ flux from 
the SN 1987A. 
In particular, these data indicate that photon fluence 
for for the photon energy band $E_\gamma=4.1-6.4$ MeV 
integrated over first 10 s from the collapse, 
did not exceed one photon per cm$^{2}$. 
Therefore, the only way to to suppress the gamma 
ray signal from  is to assume that the dominant 
part of the axions have undergone the decay inside the 
star envelope, during the flight time $t_s=cR_s \sim 100$ s,  
where $R_s \sim 3\cdot 10^7$ km is a radius of 
the SN 1987A progenitor star, the blue giant Sanduleak.     
Hence, for the relevant axion energies 
$E\simeq 2E_\gamma =8-13$ MeV, the exponential factor 
$\exp(-m_a \Gamma_{\rm tot}t_s/E)$ should be very small, 
where the total decay width 
$\Gamma_{\rm tot}$ incorporates the decays 
into the ordinary and mirror photons 
(\ref{width}), and if $m_a >2m_e$, 
also into electron-positron (\ref{width-e}). 
Therefore, $\Gamma_{\rm tot}^{-1}$ should not exceed 
few seconds. 

More precisely, by taking that the axions emitted from 
the axiosphere have a thermal spectrum with a 
temperature which at the time moment $t$ from the collapse 
estimated as 
$T_a \simeq f_6^{4/11} (t/10 ~{\rm ms})^{-1/4}\times 6$ MeV 
\cite{Bur}, one can compute the expected photon 
fluence in the energy band $E_\gamma=4.1-6.4$ MeV and 
confront it to the SMM bound. The results for various 
values of the parameter $x$ are shown in Fig. 2. 
Here the dash curves correspond to the SMM limit on 
the $\gamma$ fluence and so they set an upper limit on $f_a$ 
for given values of $x$ and $x^\prime$.  
 
Thus, we see that there is a parameter space 
compatible with the existing experimental and astrophysical 
limits, which yields $f_a$ around $10^4-10^5$ GeV 
and $m_a$ about 1 MeV. This parameter range can be 
accessible for the experimental testing at the 
future reactor or beam dump experiments.

\section{Planck scale corrections}

A potential problem for the invisible axion models 
arises from the Planck scale effects. 
Many arguments suggest that 
the non-perturbative quantum gravity effects 
can induce the higher order effective
operators cutoff by the Planck scale $M_{P}$ 
which explicitly violate the global symmetry. 
Let us take the DFSZ model \cite{DFSZ} in which 
$U(1)_{\rm PQ}$ symmetry is spontaneoulsy broken by 
the gauge singlet scalar $S$, $\langle S\rangle=f_a/\sqrt2$    
and consider a dimension 5 operator  
$(\lambda S(S^\dagger S)^2/M_{P} + \mathrm{h.c.}$,  
with $\lambda$ being some complex coupling constant.   
This term explicitly breaks the PQ symmetry and induces 
the mass term to the axion field: 
\begin{equation}  \label{M-a}
M_a^2 = \frac{\lambda f_a^3}{2\sqrt2 M_P} \simeq 
\lambda f_6^3 \times 0.17 ~{\mathrm{GeV}}^2
\end{equation}
which should be compared with the 
dynamical mass term $m_a^2$ (\ref{a-m}).
The potential for $\bar\theta = a/f_a$ then becomes 
$f_a^{-2} P(a)=m_{a}^2 \cos \bar\theta + 
M_a^2 \cos (\bar\theta + \delta)$, 
where $\delta$ is related to the complex phase of the coupling 
constant $\lambda$ and it is generically order 1. 
Although the first "genuine" term in this potential 
would fix the axion VEV at $\langle \bar\theta \rangle = 0$, 
the second term does not care about the strong CP problem 
and tends to minimize
the potential along $\langle \bar\theta \rangle = -\delta$. 
Therefore, $\theta < 10^{-10}$ demands that 
$M_a <10^{-5} m_a$,   
which in turn translates into the upper limit 
$f_a < \lambda^{-1/5} \times 10$ GeV.  
This is in sharp contradiction with the lower limit 
$f_a > 10^{10}$ GeV unless the constant 
$\lambda$ is extremely small, $\lambda < 10^{-45}$.
Therefore, if one takes the Planck scale induced effects 
seriously, there is no room left for the invisible axion 
models.

As for our model, it appears to be stable against the 
Planck scale corrections by threefold reason: (i) 
the PQ scale is lower, $f_a = 10^4-10^5$ GeV,  
(ii) the axion is heavier, $m_a\sim 1$ MeV, and  
(iii) the minimal possible explicit Planck scale operators 
in the Higgs potential are of dimension 6  
and so they are suppressed $M_{P}^2$: 
Let us consider, e.g. the terms:  
\begin{equation}
\lambda \frac{(H_1 H_2)^3}{M_P^2} +\lambda 
\frac{(H^{\prime}_1 H^{\prime}_2)^3}{M_P^2} ~ + ~{\rm h.c.}  
\label{dim-6}
\end{equation}
They induce the axion explicit mass term: 
\begin{equation}
M_a^2 = 
\frac{9\lambda(x^\prime+x^{\prime -1}) f_a^4}{4M_{P}^2} 
\simeq  \lambda x^\prime f_6^4 \times 0.5\cdot {\rm keV}^2
\label{h-dim}
\end{equation}
and the constraint $M_a <10^{-5} m_a$,  
for a typical value $m_{a} \sim 1$ MeV, 
implies $f_a < 1.2 (\lambda x^\prime)^{-1/4} 10^5$ GeV, 
which is consistent with the parameter range 
found in previous section.  
 
The Planck scale effects could explicitly break the PQ symmetry 
also in the Yukawa like operators \cite{BMS}, e.g.  
\begin{equation}  \label{Pl-Yuk} 
a^{ij} \frac{H_1 H_2}{M_P^2} d^c_i q_j H_1 +
a^{\prime ij} \frac{H'_1 H'_2}{M_P^2} d^{c\prime}_i q'_j H'_1
\end{equation}
which would interfere with the contribution to the Yukawa terms 
(\ref{SM-Yuk}) and could deviate $\bar{\theta}$ from zero. 
However, for $f_a < 10^6$ GeV these corrections are also 
negligible and do not give problems.

Let us now comment about the model of Rubakov \cite{Rub} 
which was based on the grand unified 
picture $SU(5)\times SU(5)'$. It was assumed that the 
mirror GUT symmetry $SU(5)'$ breaks down to the 
mirror standard model gauge group 
$SU(3)' \times SU(2)' \times U(1)'$ at lower scale 
than the ordinary $SU(5)$ breaks to standard gauge group. 
In other words, the VEV of the Higgs 24-plet $\Phi'$ of 
$SU(5)'$ is somewhat lower than the VEV of the 24-plet Higgs 
of $SU(5)$, 
$\langle \Phi \rangle > \langle \Phi' \rangle$. 
Then the difference between the infrared poles 
for the strong coupling constants $\Lambda'$ and $\Lambda$  
can emerge obtained due to the RG evolution 
from the corresponding scales down in energy. 
The gauge constant of $SU(5)$ runs faster then the one of $SU(3)$, 
and thus it is possible to have the bigger QCD scale 
in mirror sector, $\Lambda'\gg \Lambda$.  
In this way one could obtain obtain a very massive axion 
(e.g. $m_a \sim 100$ GeV or more). 

However, in this case the M-parity is broken at GUT scales 
and therefore there is no strong reason to expect that 
it will be valid in the low energy Yukawa sector. 
For example, the Planck scale operators of dimension 5 
are dangerous: 
\begin{equation} \label{rub-P}
G \overline{5} 10 \overline{H}+ 
G' \overline{5^{\prime}} 10^{\prime}\overline{H^{\prime}} + 
k \frac{\Phi}{M_P} \overline{5} 10 \overline{H} + 
k' \frac{\Phi^{\prime}}{M_P} 
\overline{5^{\prime}} 10^{\prime}\overline{H^{\prime}}... 
\end{equation}
where $H$ is the 5-plet Higgs. 
Due to M-parity we have $G=G'$ and $k=k'$ but the
effective Yukawa coupling 
$\overline{G}=G+k \langle \Phi \rangle /M_P$ is
not equal to the mirror effective Yukawa coupling 
$\overline{G'}=G+k\langle \Phi' \rangle /M_P$. 
This should cause the big phase difference beteween them 
and so that $\overline{\theta'} \neq \overline{\theta}$.
Therefore, the model \cite{Rub} is not stable against 
the Planck scale corrections. 

A way to overcome this problem is to break M-parity only at lower 
energies. E.g., one can take $SU(5)\times SU(5)'$ model  
with $\langle \Phi \rangle = \langle \Phi' \rangle$,  
and break M-parity only 
due to different electroweak scales in two sectors, 
so that  $\overline{G'}=\overline{G'}$. 
Clealy, this is nothing but embedding of our model 
considered in this paper in the grand unified picture.

\section{Conclusions}

We have suggested a new possibility for solving 
the strong CP-problem. 
It assumes that in parallel to the observable particle world 
described by the standard model, 
there exists a mirror world of particles with the identical 
Lagrangian, and two sectors 
share the same Peccei-Quinn symmetry realized {\it \`a la} 
Weiberg-Wilczek model. 
We asume that the mirror symmetry between two worlds 
is spontaneously broken so that the electroweak scale 
in the mirror sector $v^\prime$ is considerably larger 
than the ordinary electroweak scale $v=247$ GeV. 
This in turn implies the infrared scale $\Lambda'$ of the mirror 
strong interactions has to be somewhat larger than the 
ordinary QCD scale $\Lambda\simeq 200$ MeV. 
In this way, the axion mass and interaction constants 
are actually determined by mirror sector scales $v'$ 
and $\Lambda'$, while the $\theta$ terms are 
simultaneously cancelled in both sectors due to mirror 
symmetry. 

This means that the axion state essentially emerges from the 
mirror Higgs doublets, up to small ($\sim v/v^{\prime}$) 
admixture from the ordinary Higgses. 
Hence, our axion is a WW-type axion for the mirror sector 
while from the point of view of ordinary sector it is 
DFSZ-type invisible axion with the symmetry breaking scale 
$f_a \sim v^{\prime}\gg v$.

We have discussed the experimental and astrophysical limits on 
such an axion and have found an interesting parameter 
window where the PQ scale is $f_a \sim 10^4-10^5$ GeV and 
the axion mass is $m_a \sim 1$ MeV, 
which is accessible for the axion search in the future 
reactor and beam dump experiments.  
Interestingly, for such a parameter range our model 
is stable against the Planck scale corrections as far as 
the strong CP-problem is concerned. 

In our model, which in fact is the simplest mirror extension 
of the standard model, we cannot approach parameter 
space which could be applicable for explanation of 
the Gamma Ray Bursts via axion emission from the collapsing 
compact objects \cite{axidragon}. 
This, however, could be easily achieved 
in the supersymmetric version of our model, due to 
stronger effects in the renormalization group evolution of 
the strong coupling constants  in two sectors.

\vspace{3mm}
{\large \textbf{Acknowledgements}} 
\vspace{3mm}

We thank Anna Rossi who participated an earlier stage 
of this work. 
The work is partially supported by the MURST research grant 
``Astroparticle Physics".


\begin{thebibliography}{99}
\bibitem{kim}  J.E. Kim, Phys. Rep. 150 (1987) 1; 
H.Y. Cheng, \textit{ibid.} 158 (1988) 1.

\bibitem{PQ}  R.D. Peccei, H.R. Quinn, Phys. Rev. D 16 (1977) 1791. 

\bibitem{WW}  S. Weinberg, Phys. Rev. Lett. 40 (1978) 223; F. Wilczek, 
\textit{ibid.} 40 (1978) 279.

\bibitem{DFSZ}  A.R. Zhitnitskii, Sov. J. Nucl. Phys. 31 (1980) 260; 
M. Dine, W. Fischler, M. Srednicki, Phys. Lett. B104 (1981) 199.

\bibitem{KSVZ}  J.E. Kim, Phys. Rev. Lett. 43 (1979) 103; 
M. Shifman, A. Vainshtein, V. Zakharov, Nucl. Phys. B166 (1980) 493.

\bibitem{archion}  Z. Berezhiani, Phys. Lett. B129 (1983) 99; 
Phys. Lett. B150 (1985) 177; 
Z. Berezhiani, M. Khlopov, Z. Phys. C 49 (1991) 73; Sov. J.
Nucl. Phys. 51 (1990) 739; \textit{ibid.} 51 (1990) 935.

\bibitem{Choi}  
K. Choi, K. Kang, J.E. Kim, Phys. Lett. B181 (1986) 145.

\bibitem{PDG}  Particle Data Group, Eur. Phys. J. C15 (2000) 1.

\bibitem{raffelt}  G.G. Raffelt, Phys. Rep. 198 (1990) 1.

\bibitem{kamion}  
R. Holman \textit{et al.}, Phys. Lett. B282 (1992) 132; 
M. Kamionkowski, J. March-Russel, Phys. Lett. B282 (1992) 137; 
S. Barr, D. Seckel, Phys. Rev. D46 (1992) 539; 
S. Ghigna, M. Lusignoli, M. Roncadelli, Phys. Lett. B283 (1992) 278; 
E. Akhmedov \textit{et al.,} Phys. Lett. B 299 (1993) 90.

\bibitem{Parity}  
A. Nelson, Phys. Lett. B136 (1983) 387; 
S. Barr, Phys. Rev. Lett. 53 (1984) 329; 
K.S. Babu, R.N. Mohapatra, Phys. Rev. D 41 (1990) 1286; 
S. Barr, D. Chang, G. Senjanovi\'c, Phys. Rev. Lett. 67 (1991) 2765; 
Z. Berezhiani, Mod. Phys. Lett. A 6 (1991) 2437.

\bibitem{BMS}  Z. Berezhiani, R.N. Mohapatra, G. Senjanovi\'c, 
Phys. Rev. D47 (1993) 5565.

\bibitem{mirror}
T.D. Li, C.N. Yang, Phys. Rev. 104 (1956) 254; 
Y. Kobzarev, L. Okun, I. Pomeranchuk, Yad. Fiz. 3 (1966) 1154; 
S. Blinnikov, M. Khlopov, Astron. Zh. 60 (1983) 632; 
B. Holdom, Phys. Lett. B166 (1985) 196; 
S.L. Glashow, Phys. Lett. B167 (1986) 35; 
M. Khlopov {\it et al.}, Astron. Zh. 68 (1991) 42; 
R. Foot, H. Lew, R. Volkas, Phys. Lett. B272 (1991) 67; 
Mod. Phys. Lett. A9 (1994) 169; 
R. Foot, R. Volkas, Phys. Rev. D 52 (1995) 6595; 
Z. Silagadze, Phys. At. Nucl. 60 (1997) 272; 
S. Blinnikov, astro-ph/9801015, astro-ph/9902305; 
Z. Berezhiani, D. Comelli, F. Villante, hep-ph/0008105, 
and references therein. 

\bibitem{PLB98}  
Z. Berezhiani, Phys. Lett. B417 (1998) 287.

\bibitem{BM}  
Z. Berezhiani, R.N. Mohapatra, Phys. Rev. D 52 (1995) 6607; 
see also E. Akhmedov, Z. Berezhiani, G. Senjanovi\'c, 
Phys. Rev. Lett. 69 (1992) 3013.

\bibitem{BDM}  
Z. Berezhiani, A. Dolgov, R.N. Mohapatra, Phys. Lett. B375 (1996) 26; 
Z. Berezhiani, Acta Phys. Polon. B 27 (1996) 1503; 
R.N. Mohapatra, V. Teplitz, Astrophys. J. 478 (1997) 29; 
Phys. Lett. B462 (1999) 302; astro-ph/0001362; 
V. Berezinsky, A. Vilenkin, Phys. Rev. D 62 (2000) 083512. 

\bibitem{Rub}  
V.A. Rubakov, JETP Lett. 65 (1997) 621. 

\bibitem{Parida}
D. Chang, R. Mohapatra, M. Parida, Phys. Rev. D 30 (1984) 1052. 

\bibitem{Yulik}  
J. Ellis, M.K. Gaillard, Nucl. Phys. B150 (1979) 141; 
I.B. Khriplovich, Phys. Lett. B173 (1986) 193; 
I.B. Khriplovich and A.I. Vainshtein, Nucl. Phys. B414 (1994) 27.

\bibitem{altmann}  M.~Altmann \textit{et al.,} 
Z.\ Phys.\ C68 (1995) 221.

\bibitem{faissner}  
H.~Faissner \textit{et al.}, Z.\ Phys.\ C37 (1988) 231.

\bibitem{Tur}  M.S. Turner, Phys. Rev. Lett. 60 (1988) 1797.

\bibitem{Bur}  
A. Burrows, M.T. Ressel, M.S. Turner, Phys. Rev. D 42 (1990) 3297.

\bibitem{SMM} 
E. Chupp, W. Westrand, C. Reppin, Phys. Rev. Lett. 62 (1989) 505; 
E.W. Kolb, M.S. Turner, Phys. Rev. Lett. 62 (1989) 509.

\bibitem{axidragon} 
Z. Berezhiani, A. Drago, Phys. Lett. B473 (2000) 281, hep-ph/9911333.

\end{thebibliography}
\end{document}